\RequirePackage{lineno}
\documentclass[12pt,twoside]{article}   
\usepackage[super,sort,comma]{natbib}

\usepackage{fancyhdr}		

\usepackage[section]{placeins}   %

\usepackage{url}
\usepackage{caption} 
\usepackage{framed,multirow}

\usepackage{amssymb}
\usepackage{amsmath}
\usepackage{latexsym}

\usepackage{color, soul}
\usepackage{xargs}
\usepackage{tikz}
\usepackage{todonotes}
\usepackage{multirow} 
\usepackage{verbatim}
\usepackage{nameref,hyperref}

\usepackage{url}
\usepackage{xcolor}

\usepackage{hyperref}

\usepackage{graphicx}

\makeatletter \renewcommand\@biblabel[1]{$^{#1}$} \makeatother
\usepackage{hyperref}
\hypersetup{ colorlinks,
    citecolor=blue,
    filecolor=blue,
    linkcolor=blue,
    urlcolor=blue
}

\usepackage{xcolor}

\definecolor{gray}{rgb}{0.6,0.6,0.6}
\definecolor{red}{rgb}{0.85,0,0}
\definecolor{green}{rgb}{0,0.85,0}
\definecolor{blue}{rgb}{0,0,0.85}
\definecolor{beige}{rgb}{0.92,0.87,0.78}
\usepackage[all]{hypcap}    

\begin{document}

\Large {\bfseries Deep cross-modality (MR-CT) educed distillation learning for cone beam CT lung tumor segmentation } \\  
\vspace*{10mm}
\\
\small{Jue Jiang$^{1}$,Sadegh Riyahi Alam$^{1}$,Ishita Chen$^{2}$,Perry Zhang$^{1}$,Andreas Rimner$^{2}$ \\ Joseph O. Deasy$^{1}$,Harini Veeraraghavan$^{1}$}\\		
Department of Medical Physics$^{1}$ \\
Department of Radiation Oncology$^{2}$ \\
Address: Box 84 - Medical Physics, Memorial Sloan Kettering Cancer Center, 1275 York Avenue, New York, NY 10065.\\
email: veerarah@mskcc.org \\	

\pagenumbering{roman}
\setcounter{page}{1}
\pagestyle{plain}


\begin{abstract}
\noindent {\bf Purpose:} Despite the widespread availability of in-treatment room cone beam computed tomography (CBCT) imaging, due to the lack of reliable segmentation methods, CBCT is only used for gross set up corrections in lung radiotherapies. Accurate and reliable auto-segmentation tools could potentiate \textcolor{black}{volumetric response assessment} and geometry-guided adaptive radiation therapies. Therefore, we developed a new deep learning CBCT lung tumor segmentation method. \\
{\bf Methods:} The key idea of our approach called cross modality educed distillation (CMEDL) is to use magnetic resonance imaging (MRI) to guide a CBCT segmentation network training to extract more informative features during training. We accomplish this by training an end-to-end network comprised of unpaired domain adaptation (UDA) and cross-domain segmentation distillation networks (SDN) using unpaired CBCT and MRI datasets. \textcolor{black}{UDA approach uses} CBCT and MRI that are not aligned and may arise from different sets of patients. The UDA network synthesizes pseudo MRI from CBCT images. The SDN consists of teacher MRI and student CBCT segmentation networks. \textcolor{black}{Feature distillation regularizes the student network to extract CBCT features that match the statistical distribution of MRI features extracted by the teacher network and obtain better differentiation of tumor from background.} The UDA network was implemented with a cycleGAN improved with contextual losses. We evaluated both Unet and dense fully convolutional segmentation networks (DenseFCN). Performance comparisons were done against CBCT only \textcolor{black}{using 2D and 3D} networks. We also compared against an alternative framework that used UDA with MR segmentation network, whereby segmentation was done on the synthesized pseudo MRI representation. \textcolor{black}{All networks were trained with 216 weekly CBCTs and 82 T2-weighted turbo spin echo MRI acquired from different patient cohorts. Validation was done on 20 weekly CBCTs from patients not used in training. Independent testing was done on 38 weekly CBCTs from patients not used in training or validation.} Segmentation accuracy was measured using surface Dice similarity coefficient (SDSC) and Hausdroff distance at 95th percentile (HD95) metrics.\\
{\bf Results:} The CMEDL approach significantly improved (p $\textless$ 0.001) the accuracy of both Unet (SDSC of 0.83 $\pm$ 0.08; HD95 of 7.69 $\pm$ 7.86mm) and DenseFCN (SDSC of 0.75 $\pm$ 0.13; HD95 of 11.42 $\pm$ 9.87mm) over CBCT only \textcolor{black}{2D}Unet (SDSC of 0.69 $\pm$ 0.11; HD95 of 21.70 $\pm$ 16.34mm), \textcolor{black}{3D Unet (SDSC of 0.72$\pm$ 0.20; HD95 15.01$\pm$12.98mm)}, and DenseFCN (SDSC of 0.66 $\pm$ 0.15; HD95 of 22.15 $\pm$ 17.19mm) networks. The alternate framework using UDA with the MRI network was also more accurate than the CBCT only methods but less accurate the CMEDL approach. \\
{\bf Conclusions:} Our results demonstrate that the introduced CMEDL approach produces reasonably accurate lung cancer segmentation from CBCT images. Further validation on larger datasets is necessary for clinical translation. 
\end{abstract}

\newpage     

\pagenumbering{arabic}
\setcounter{page}{1}
\pagestyle{fancy}

	\section{Introduction}
    Lung cancer is the leading cause of cancer-related deaths in both men and women in the United States\citep{seigel2020}. The standard treatment for inoperable or unresectable stage III \textcolor{black}{\textcolor{black}{locally advanced non-small cell lung cancer} (LA-NSCLC)} cancers is definitive/curative radiotherapy to 60 Gy in 30 fractions with concomittant chemotherapy\citep{bradley2020}. Recently, dose escalation trials and high-dose adaptive radiotherapies\citep{Weiss2013,Kavanaugh2019} have shown the feasibility to improve local control and survival in LA-NSCLC. However, a key technical challenge in delivering high-dose treatments, both at the time of planning and delivery is accurate and precise delineation of both target tumors and normal organs\citep{Sonke2010}.    
    
    Importantly, X-ray cone beam CT (CBCT) imaging is available as part of standard equipment. However, much of the in-treatment-room CBCT information cannot be used routinely beyond basic positioning corrections. In fact, a key obstacle for clinical adoption of adaptive radiotherapy for LA-NSCLC is the lack of reliable segmentation tools, needed for geometric corrections in target and possibly the critical OARs\citep{Kavanaugh2019,Sonke2019}. Despite the widespread development of deep learning segmentation methods, to our best knowledge, there are no reliable CBCT methods for routine lung cancer treatment. The latest works published were primarily focused on the pelvic regions for prostate cancer radiotherapy\citep{jia2019,Fu2020,Lei2020}.
    
    The difficulty in generating accurate segmentation results from the lack of sufficient soft-tissue contrast on CBCT imaging, especially for centrally located cancers. Low soft-tissue contrast makes it inherently difficult to extract features that clearly differentiate the target from its background structures even for a deep learning method. Prior works\citep{Fu2020,Lei2020} have used pseudo MRI (pMRI) produced from CBCT to produce more accurate pelvic organs segmentation than CBCT alone. The key idea is that pMRI, which mimics the statistical intensity characteristics of MRI contains better soft tissue contrast than CBCT, which helps accuracy. Our approach improves this idea whereby MRI is used to regularize the extraction of more informative CBCT segmentation features even on less informative CBCT modality. Unlike the approach in Fu et.al\citep{Fu2020}, which required paired CT and MRI image sets, our approach uses unpaired CBCT and MRI images, which are easier to obtain and practically applicable without requiring specialized imaging protocols for algorithm development. 
    
    Our approach introduces unpaired cross-modality distillation learning. Distillation learning was introduced to compress the knowledge contained in an information rich, high-capacity network (trained with a large training data) into a small, low-capacity network\citep{hinton2015distilling}, using paired images. Model compression is meaningful when a high-capacity model is not required for a task or the computational limitation of using such a large classifier necessitates the use of simpler and computationally fast model \citep{bucilua2006model} for real-time analysis. The approach in\citep{hinton2015distilling} used same modality images and was used to solve different image-based classification tasks. Distillation itself was done by using the probabilistic ''softMax" outputs of the teacher as target output for the student (compressed) network. Improvements to this approach included hint learning for image classification, where features from intermediate layers in the student network are constrained to mimic the features from the teacher network \citep{romero2014fitnets,li2017mimicking}. Recent works in computer vision, extended this approach using low- and high-resolution images\citep{su2016cross}, as well as for different modality distillation between paired (Red, Green, Blue) or RGB and depth images\citep{gupta2016cross} for semantic segmentation. Our work extends this approach to unpaired distillation learning using explicit hints between MRI and CBCT images. Also, we modify how the knowledge distillation is employed, whereby instead of transferring knowledge from a very deep network into a smaller network, the teacher and student networks are identical except for the imaging modalities used to train them. The teacher network is trained with a more informative MRI, \textcolor{black}{while the student CBCT network is \textcolor{black}{regularized} to extract similar features for inference like the teacher network.} As a result, only the CBCT segmentation network is required for testing. Methods that require pMRI as an additional input\citep{Fu2020,jia2019} need both cross-modality I2I translation and the segmentation network at testing time.  
    
    
    This work builds on our prior work that used unpaired MRI and contrast enhanced CT (CECT) datasets to improve CECT lung tumor segmentation\citep{jiang2019}. Our approach called cross-modality educed distillation learning (or CMEDL) extends this approach to more challenging CBCT images. We tested the hypothesis that MRI information extracted using unpaired CBCT and MRI can \textcolor{black}{regularize features} computed by the CBCT network and improve performance over CBCT only segmentation. End-to-end network training also benefits from these losses to improve I2I translation. 
 
    Our contributions include: (i) a new unpaired cross-modality educed distillation-based segmentation framework for regularizing inference on less informative modality by using more informative imaging modality, (ii) application of this framework to the challenging CBCT lung tumor segmentation, and (iii) implementation of our framework using two different segmentation networks, with performance comparisons done against other related approaches.  
    
    \section{Materials and Methods}
	\subsection{Patient and image characteristics}
	\textcolor{black}{A total of 274 weekly CBCT scans from 69 unique patients diagnosed with LA-NSCLC and treated with conventionally fractionated radiotherapy, and sourced from 49 internal and 20 external institution dataset\citep{hugo2017} were analyzed. The internal scans had segmentations on weekly CBCTs with a maximum of 7 per patient. \textcolor{black}{Only week 1 CBCTs from the external dataset were analyzed.} \textcolor{black}{Two of 49 internal patients had seven weekly CBCTs; 18 had six weekly  CBCTs; 18 had five weekly CBCTs; 9 had four weekly CBCTs and the remaining 2 had three weeks CBCT segmented.}}
	
	The internal CBCT scans were acquired for routinely monitoring geometric changes of tumor in response to radiotherapy. The external institution 4D CBCT scans were orginally collected for investigating breathing patterns of LA-NSCLC patients undergoing chemoradiotherapy. Image resolution for the weekly 4DCBCT were 0.98 to 1.17 mm in-plane spacing and 3mm slice thickness. 
	
	Each CBCT image was standardized and normalized using its global mean and standard deviation and then registered to the planning CT scans using a multi-resolution B-spline regularized diffeomorphic image registration\cite{Tustison2013, alam2020}. All contours were reviewed by a radiation oncologist and modified when necessary and served as expert delineation. B-spline registration was performed using a mesh size of 32mm at the coarsest level and was reduced by a factor of two at each sequential level. The optimization step was set to 0.2 with the number of iterations (100, 70, 30) at each level. Additional details of this registration for these datasets are in\cite{alam2020}. 
	
	 Eighty one T2-weighted turbo spin echo (TSE) MRI for cross-modality learning was obtained from 28 stage II-III LA-NSCLC patients scanned every week on a 3T Philips Ingenia scanner. \textcolor{black}{Eleven out of these 28 had weekly MRI scans ranging between 6 to 7 weeks. Seven of these 11 patients overlapped with the internal MSK CBCT cohort. However, the MRI and CBCT images were neither co-registered nor treated as paired image sets for the purpose of training.} The MRI scan parameters were: 16-element phased array anterior coil and a 44-element posterior coil (TE/TR = 120/3000-6000ms, slice thickness of 2.5mm, in-plane pixel size of 1.1 $\times$ 0.97$mm^2$, flip angle of 90$^{\deg}$, number of averages = 2, and field of view of 300 $\times$ 222 $\times$ 150$mm^3$.   
	
    \subsection{Approach}
\begin{figure*}
		\begin{center}
			\includegraphics[width=0.9\columnwidth,scale=0.5]{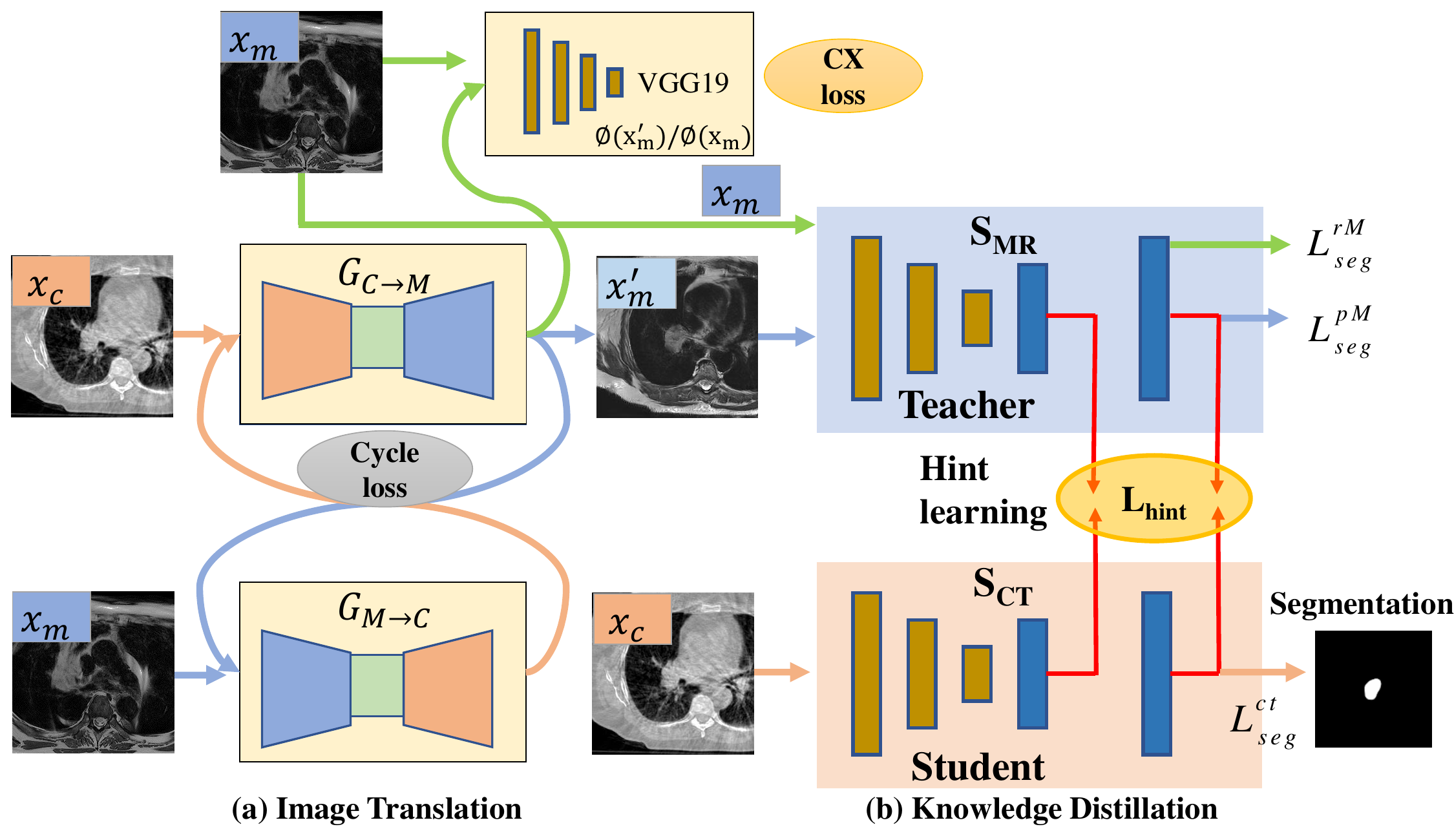}
			\vspace{-0.05cm}\setlength{\belowcaptionskip}{-0.4cm}\setlength{\abovecaptionskip}{0.08cm}\caption{\label{fig:methods} \small Approach overview. $x_{c}, x_{m}$ are the CBCT and MR images from unrelated patient sets; $G_{C \rightarrow M}$ and $G_{M \rightarrow C}$ are the CBCT and MRI translation networks; $x_{m}^{'}$ is the pseudo MRI (pMRI) image; $x_{c}^{'}$ is the pseudo CBCT image; $S_{MR}$ is the teacher network; $S_{CT}$ is the student CBCT segmentation network; CX loss is contextual loss. $L^{rM}_{seg}, L^{pM}_{seg}$ are segmentation losses used to train the teacher network, while $L^{ct}_{seg}$ is the loss for the student CBCT network.}
		\end{center}
	\end{figure*}

    An overview of our cross-modality educed distillation (CMEDL) approach is shown in Fig.~\ref{fig:methods}. The end-to-end trained network consists of an unpaired cross-domain adaptation (UDA) network composed of a generational adversarial network (GAN)\cite{goodfellow2014generative} and a segmentation distillation network (SDN), which includes a teacher MRI and student CBCT segmentation network. The teacher network is trained with expert segmented MRI and pMRI images. TheCBCT network is trained with expert-segmented CBCT images. Feature distillation is performed by hint learning\cite{romero2014fitnets}, whereby feature activations on the CBCT network in specific layers (last and penultimate) are forced to mimic the corresponding layer feature activations of the teacher network extracted from corresponding synthesized pMRI images.
    
    \subsection{Notations}
    The network is trained using a set of expert-segmented CBCT $\{x_c, y_c\} \in \{X_{C}, Y_{C}\}$ and MRI $\{x_m, y_m\} \in \{X_{M}, Y_{M}\}$ datasets. The CBCT and MRI do not have to arise from the same sets of patients and are not aligned for network training. The cross-modality adaptation network consists of generators $G_{C \rightarrow M}: x_{c} \mapsto x_{m}$ to produce pseudo MRI $x_{m}^{\prime}$, $G_{M \rightarrow C}: x_{m} \mapsto x_{c}$ to produce pseudo CT images $x_{c}^{\prime}$, and domain discriminators $D_M$ and $D_C$. The sub-networks $G_{M \rightarrow C}$ and $D_{C}$ are used to enforce cyclically consistent transformation when using unpaired CBCT and MRI datasets. Feature vectors produced through a mapping function $F(x): x \mapsto f(x)$ are indicated using italized text \textit{$f_j$\/}\rm, where $j=1, \ldots K$, for $K = H \times W \times C$ for 2D and $K = H \times W \times Z \times C$ for 3D, is the number of features for an image of height $H$, width $W$, depth $Z$, and channels $C$.  
    
    \subsection{Stage I: Unpaired cross-domain adaptation for image-to-image translation}
    The UDA network is composed of a pair of GANs for producing pseudo MRI $x_{m}^{\prime}$ and pseudo CT $x_{c}^{\prime}$ images using generator networks $G_{C \rightarrow M}$ and $G_{M \rightarrow C}$, respectively. The images produced by these generators are constrained by global intensity discriminators $D_M$ and $D_C$ for MRI and CBCT images, respectively. The adversarial loss for these two networks are computed as:
    \begin{equation}
	\begin{split}
	\setlength{\abovedisplayskip}{0pt}
	\setlength{\belowdisplayskip}{0pt}
	& L^{M}_{adv}(G_{C \rightarrow M}, D_{M}, X_{M}, X_{C})= \mathbb{E}_{x_{m} \sim X_{M}} [log(D_{M}(x_{m}))] +  \mathbb{E}_{x_{c} \sim X_{C}} [log(1-(D_{M}(G_{C \rightarrow M}(x_{c}))] \\
	& L^{C}_{adv}(G_{M \rightarrow C}, D_{C}, X_{C}, X_{M})  = \mathbb{E}_{x_{c} \sim X_{C}} [log(D_{C}(x_{c}))] + \mathbb{E}_{x_{m} \sim X_{M}} [log(1-(D_{C}(G_{M \rightarrow C}(x_{m}))]. 
	\end{split} 
	\label{eqn:adversary loss_MRI}
	\end{equation}
	
	Because the networks are trained with unpaired images, cyclical consistency is enforced to shrink the space of possible mappings computed by the generator networks. The loss to enforce this constraint is computed by minimizing the pixel-to-pixel loss (e.g. L1-norm) between the generated (e.g. $G_{ \circlearrowleft M} = G{M \rightarrow C}(G_{C \rightarrow M}(x_c))$) and original images as:
	
	\begin{equation}
	\begin{split}
	L_{cyc}(G_{C \rightarrow M}, G_{M \rightarrow C},X_{c}, X_{m})  =  \mathbb{E}_{x_{c} \sim X_{c}}\left[\left\|G_{C\circlearrowleft M}(x_{c}) - x_{c}\right\|_{1}\right] +  \mathbb{E}_{x_{m} \sim X_{m}}\left[\left\|G_{M\circlearrowleft C}(x_{m}) - x_{m}\right\|_{1}\right].
	\end{split}
	\end{equation}
    
    However, the cyclical consistency loss alone can only preserve global statistics while failing to preserve organ or target specific constraints\cite{jiang2018tumor}. Furthermore, when performing unpaired adaptation between unaligned images, pixel-to-pixel matching losses are inadequate to preserve spatial fidelity of the structures \cite{mechrez2018contextual}. Therefore, we used the contextual loss that was introduced in\cite{mechrez2018contextual}. The contextual loss is computed by matching the low- or mid-level features extracted from the generated and the target images using a pre-trained network like the VGG19\cite{simonyan2014very} (default network used in this work). This loss is computed by treating the features as a collection and by computing all feature-pair similarities, thereby, ignoring the spatial locations. In other words, the similarity between the generated ($f(G(x_{c})) = {g_{j}}$) and target feature maps ($f(x_{m})={m_{i}}$) are marginalized over all feature pairings and the maximal similarity is taken as the similarity between those two images. Therefore, this loss also considers the textural aspects of images when computing the generated to target domain matching. It is similar to perceptual losses, but ignores the spatial alignment of these images, which is advantageous when comparing non-corresponding target and source modality generated images. The contextual similarity is computed by normalizing the inverse of cosine distances between the features $g_j$ and $m_i$ as:
	\begin{equation}
	CX(g,m) = \frac{1}{N}\sum_{j} \underset{i}max CX(g_{j},m_{i}),
	\end{equation} 
	where, $N$ corresponds to the number of features. The loss is computed as:
	\begin{equation}
	L_{cx} = -log(CX(f(G(x_{c})), f(x_{m})).
	\end{equation}
	
	The pseudo MRI images produced through I2I translation from this stage are used in the distillation learning as described below. 
	
	\subsection{Stage II: Cross-modality distillation-based segmentation}
	This stage consists of a teacher (or MRI) segmentation and a student (or CBCT) segmentation network. The goal of distillation learning is to provide hints to the student network such that features extracted in specific layers of the student network match the feature activations for those same layers in the teacher network. 
	
	Both teacher ($S_M$) and student ($S_C$) networks use the same network architecture (Unet is the default architecture), but process different imaging modalities. Both networks are trained from scratch. The teacher network is trained using expert segmented T2w TSE MRI ($\{x_m, y_m \in \{X_M, Y_M\}$) and pseudo MRI datasets ($\{x_{m}^{\prime}, y_{c}\} \in \{X_{C}, Y_{C}\}$) obtained from expert-segmented CBCT datasets. The CBCT network is trained with the expert-segmented CBCT datasets. The two networks are optimized using Dice overlap loss:
\begin{equation}
	\setlength{\abovedisplayskip}{1pt}
	\setlength{\belowdisplayskip}{1pt}
	\begin{split}
	L_{seg} & = L_{seg}^{M}+L_{seg}^{C}\\ &
	=\mathbb{E}_{x_{m}\sim X_{M}}[-log P(y_{m}|S_{M}(x_{m}))] + \mathbb{E}_{x_{m}^{\prime}\sim G_{C\rightarrow M}(X_{C})}[-log P(y_{c}|S_{M}(x_{m}^{\prime}))] \\ & + \mathbb{E}_{x_{c}\sim X_{C}}[-log P(y_{c}|S_{C}(x_{c}))].
	\label{eqn:Seg}
	\end{split}
	\end{equation}
		
	Feature distillation is performed by matching the feature activations computed on the pseudo MRI using $S_M$ and the feature activations extracted on corresponding CBCT images from the $S_C$ networks. Because the features closest to the output are the most correlated with the task\citep{lin2017refinenet}, we match the features computed from the last two network layers by minimizing the L2 loss: 

	\begin{equation}
	\setlength{\abovedisplayskip}{1pt}
	\setlength{\belowdisplayskip}{1pt}
	\begin{split}
	L_{hint} & =  \sum_{i=1}^{N} \|\phi{_{C,i}}(x_{c})-\phi{_{M,i}}(G_{C\rightarrow M}(x_{c}))||^{2} 
	\label{eqn:Feature}
	\end{split}
	\end{equation}
	where $\phi_{C,i}, \phi_{M,i}$ are the $i_{th}$ layer features computed from the two networks, $N$ is the total number of features. As identical network architecture is used in both networks, the features can be matched directly without requiring an additional step to adapt the features size as shown in  \citep{romero2014fitnets}. We call this loss the hint loss.
	\\
	The total loss is expressed as:
	\begin{equation}
	\textrm{Loss} = L_{adv} + \lambda_{cyc} L_{cyc} + \lambda_{CX} L_{CX} + \lambda_{hint} L_{hint} + \lambda_{seg} L_{seg} 
	\label{eqn:Total_loss}
	\end{equation}
	where $\lambda_{cyc}$, $\lambda_{CX}$, $\lambda_{hint}$ and $\lambda_{seg}$ are the weighting coefficients for each loss.\\
	The network update alternates between the cross-modality adaptation and segmentation distillation. The network is updated with the following gradients, $-\Delta_{\theta_{G}}(L_{adv}+ \lambda_{cyc}{L_{cyc}}+ \lambda_{CX}{L_{CX}}+ \lambda_{hint}L_{hint}+ \lambda_{seg}L_{seg}$, $-\Delta_{\theta_{D}}(L_{adv})$ and $-\Delta_{\theta_{S}}(L_{hint}+L_{seg})$.	
\subsection{Implementation details}
\paragraph{Cross-modality adaptation network structure: \/}
\textcolor{black}{The UDA network architectures were constructed based on well-proven architectures as used in our prior work for CT and MRI domain adaptation\citep{jiang2018tumor,jiang2019MIC,jiang2020} for tumor and organ at risk segmentation.} The generator architectures were adopted from DCGAN \citep{radford2015unsupervised}, \textcolor{black}{which has been proven to avoid issues of mode collapse.} Specifically, the generators consisted of two stride-2 convolutions \citep{radford2015unsupervised}, 9 residual blocks \citep{He2015} and two fractionally strided convolutions with half strides. Generator network used rectified linear unit (ReLU) \citep{radford2015unsupervised} \textcolor{black}{in order to increase stability of training. Similary,  instance normalization \citep{ulyanov2017improved} as done in \cite{zhu2017unpaired} in all but the last layer, which has a \textit{tanh} activation for image generation to increase training stability.}

	\begin{figure*}
		\begin{center}
			\includegraphics[width=1\columnwidth,scale=1]{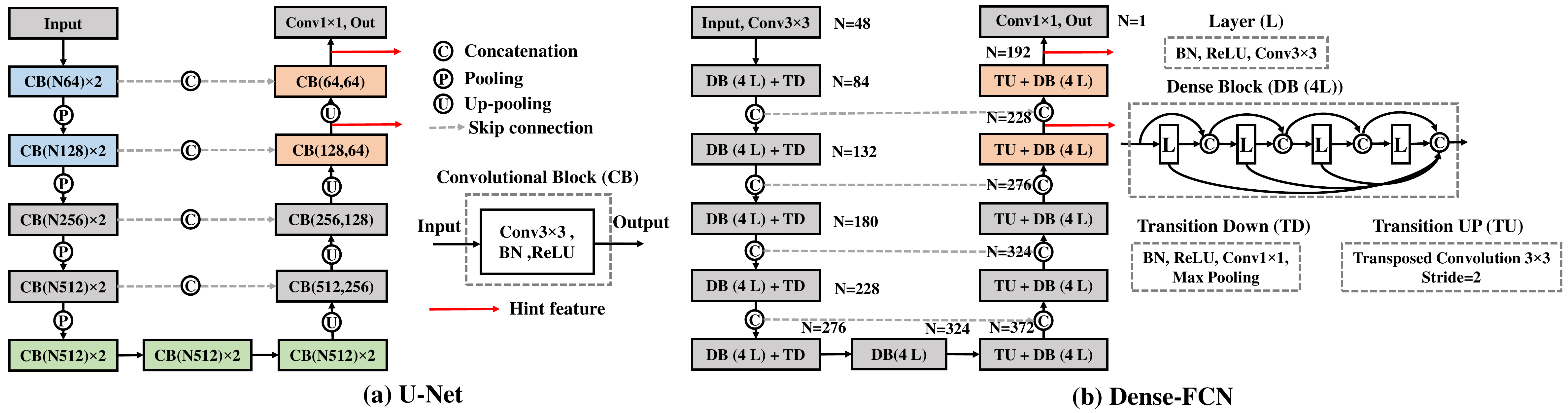}
			\vspace{-0.05cm}\setlength{\belowcaptionskip}{-0.4cm}\setlength{\abovecaptionskip}{0.08cm}\caption{\label{fig:seg_structure} \small The segmentation structure of Unet \citep{ronneberger2015u} and DenseFCN57 \citep{jegou2017one}. The red arrow indicates that the output of these layers are used for distilling information from MR into CT. This is done by minimizing the L2-norm between the features in these layers between the two networks. The blue blocks indicate the lower layer; the green blocks indicate the middle layer; the orange blocks indicate the upper layer in Unet. Best viewed in color.}
		\end{center}
	\end{figure*} 

\textcolor{black}{A patchGAN discriminator as suggested in \citep{isola2017image}, and which uses 70$\times$70 overlapping pixel image patches was used to increase the number of pixel patches to distinguish real vs. fake images to improve the stability of the discriminator. We have used a patchGAN discriminator in our prior works\citep{jiang2018tumor,jiang2020} and found it to achieve stable performance for both tumor and organ segmentation from CT and MRI. Leaky ReLU instead of ReLU was used based on the results from DCGAN\citep{radford2015unsupervised}, along with batch normalization \citep{ioffe2015batch} in all except the first and last layers to increase discriminator stability in training.}
\\
\textcolor{black}{A pre-trained VGG19 based on the standard VGGNet\citep{simonyan2014very} was used for memory efficiency.} The VGG19 consists of 16 layers of convolution filter with size 3$\times$3, 3 layers of fully connection layers and 5 maxpool layers. The lower level feature convolution filters which were of size 3$\times$3$\times$64, were progressively doubled to increase the number of feature channels while reducing the feature size through subsampling using \textit{maxpool} operation.
	\paragraph{Segmentation networks structure:}
	We implemented Unet \citep{ronneberger2015u} and DenseFCN \citep{jegou2017one} networks. \textcolor{black}{We chose these networks as these are the most commonly used segmentation architectures in medical image segmentation and have shown fairly good performance for multiple disease sites.}
	\\ 
	The U-Net \citep{ronneberger2015u} is composed of series of convolutional blocks, with each block consisting of convolution, batch normalization and ReLU activation. Skip connections are implemented to concatenate high-level and lower level features. Max-pooling layers and up-pooling layers are used to down-sample and up-sample feature resolution size. We use 4 max-pooling and 4 up-pooling in the implemented U-Net structure. The layers from the last two block of Unet with feature size of 128$\times$128$\times$64 and 256$\times$256$\times$64 are used to tie the features, shown as red arrow in Fig. \ref{fig:seg_structure} (a). This network had 13.39 M parameters and 33 layers\footnote{layers are only counted on layers that have tunable weights} of Unet. 
	
	The Dense-FCN \citep{jegou2017one} is composed of Dense Blocks (DB) \citep{huang2017densely}, which successively concatenates feature maps computed from previous layers, thereby increasing the size of the feature maps. A dense block is produced by iterative concatenation of previous layer feature maps within that block, where a layer is composed of a Batch Normalization, ReLU and 3$\times$3 convolution operation, as shown in Fig.4 (b). Such a connection also enables the network to implement an implicit dense supervision to better train the features required for the analysis.  Transition Down (TD) and Transition UP (TU) are used for down-sampling and up-sampling the feature size, respectively, where TD is composed of Batch Normalization, ReLU, 1$\times$1 convolution, 2$\times$2 max-pooling while TU is composed of 3$\times$3 transposed convolution. We use the DenseFCN57 layer structure \citep{jegou2017one}, that uses dense blocks with 4 layers for feature concatenation and 5 TD for feature down-sampling and 5 TU for feature up-sampling with a growing rate of 12.  \textcolor{black}{Although the authors of DenseFCN\citep{jegou2017one} provide implementations for deeper networks, including DenseFCN67, DenseFCN120, we used DenseFCN57 as it has the least cost when combined with the cross-modality adaptation network using the CycleGan framework.}  This resulted in 1.37 M parameters and 106 layers of DenseFCN.
\paragraph{Networks training:}
All networks were implemented using the Pytorch \citep{paszke2017automatic} library and trained end to end on Tesla V100 with 16 GB memory and a batch size of 2. The ADAM algorithm  \citep{kingma2014adam} with an initial learning rate of 1e-4 was used during training for the image translation networks. The segmentation networks were trained with a learning rate of 2e-4. We set $\lambda_{adv}$=1, $\lambda_{cyc}$=10, $\lambda_{CX}$=1, $\lambda_{hint}$=1 and $\lambda_{seg}$=5 for the coeffcient of equation \ref{eqn:Total_loss}.
	
A pre-trained VGG19 network using the ImageNet dataset was used to compute the contextual loss. The low level features extracted using VGG19 quantify edge and textural characteristics of images. Although such features could be more useful for quantifying the textural differences between the activation maps, the substantial memory requirement for extracting these features precluded their use in this work. Instead we used higher level features computed from layers Conv7, Conv8, and Conv9 that capture the mid- and high-level contextual information between the various organ structures. The feature sizes were 64$\times$64$\times$256, 64$\times$64$\times$256 and 32$\times$32$\times$512, respectively. 

\textcolor{black}{Two hundred and sixteen CBCT scans composed of 206 weekly scans from 40 internal patient, and 10 external dataset\citep{hugo2017} were used for network training. In order to increase the number of training examples and obtain a more generalizable model, the networks were trained with 42,740 2D image slices containing the tumor after cropping the original images (512 $\times$ 512) into 256 $\times$ 256 image patches of CBCT and 21,967 2D image slices of MRI image. Image cropping was done by automatic removal of regions outside the body region through intensity thresholding, followed by hole filling, and connected components extraction to identify the largest component or body region. Online data augmentation including horizontal flipping, scaling, rotation, and elastic deformation was used. Early stopping strategy was used to avoid over-fitting and the networks were trained up to utmost 100 epochs.} 

\textcolor{black}{The trained models were validated on an independent set of 20 CBCT scans arising  from 3 internal patients with 15 weekly segmented CBCTs and 5 week 1 CBCT from the external institution dataset.} \textcolor{black}{Testing was done on 38 CBCT scans from 6 internal patients with 33 weekly segmented CBCTs and 5 week 1 CBCT from the external institution dataset. Image sets from patients were separated such that all CBCT scans pertaining to a patient did not overlap across the training, validation, and testing sets to prevent any potential for data leak.}

In order to support reproducible research, we will make the code for our approach available with reasonable request upon acceptance for publication.

\section{Experiments and Results}
Experiments were done to test the hypothesis \textcolor{black}{that guiding the CBCT network training to extract as informative features as the MRI network will produce more accurate tumor segmentation on CBCT. In other words, the CBCT network is regularized to extract features similar to those extracted by the teacher network.}

We tested our distillation learning framework on two different commonly used segmentation networks, the Unet and denseFCN to measure performance differences due to network architecture. We also evaluated whether the pMRI images produced through the UDA network were more useful than the CBCT images for segmentation. For this purpose, we used the UDA and the teacher network of the CMEDL architecture. The default CMEDL network only requires the student CBCT network during testing. We also evaluated the performance of pMRI based segmentation using standard cycleGAN and a Unet segmentator, which is somewhat similar to the work in Lei et.al\citep{Lei2020} and the more advanced variational auto-encoder using the unpaired image to image translation (UNIT)\citep{li2017universal} with the Unet network. \textcolor{black}{Additionally, we evaluated whether combining the pMRI with the CBCT as an additional channel in the input resulted in performance improvement. For this experiment, the pMRI was generated using a cycleGAN . Finally, we benchmarked the performance of the 2D CMEDL network against a 3DUnet network\cite{cciccek20163d}}. 
\\
All networks were trained from scratch using identical sets of training and testing datasets. Reasonable hyper-parameter optimization was done to ensure good performance by all networks. 

\subsection{Network training stability}
\begin{figure*}
				\begin{center}	\includegraphics[width=0.9\columnwidth,scale=0.5]{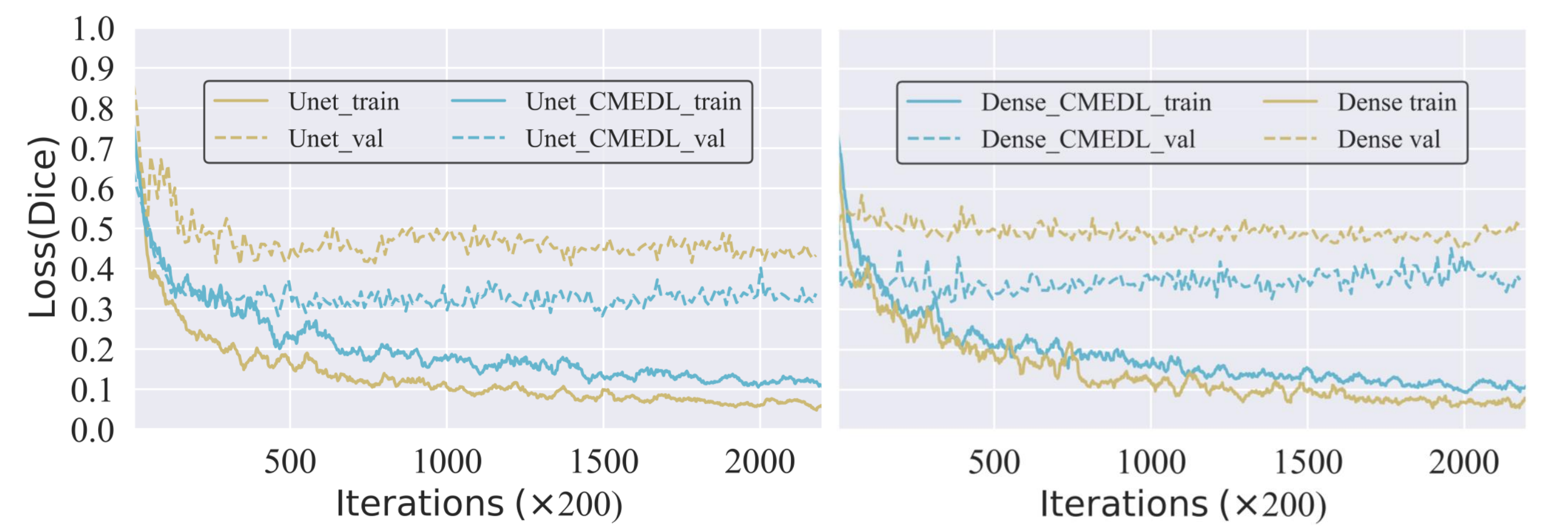}
					\vspace{-0.05cm}\setlength{\belowcaptionskip}{-0.4cm}\setlength{\abovecaptionskip}{0.08cm}\caption{\label{fig:train_curve} \small \textcolor{black}{The training and validation loss (1.0 - DSC) curves for Unet and DenseFCN networks trained with and without CMEDL approach. }}
					
				\end{center}
			\end{figure*}

\textcolor{black}{Fig.\ref{fig:train_curve} shows the training and validation loss curves for both Unet and denseFCN networks trained with and without CMEDL approach. As shown, all networks achieved stable loss performance, albeit the CMEDL approach resulted in better loss performance with the same length of training as the CBCT only network architectures.} 

\subsection{Evaluation Metrics}
Segmentation performance was evaluated from 3D volumetric segmentations produced by combining segmentations from 256$\times$256 pixel 2D image patches. In order to establish the clinical utility of the developed method, we computed surface Dice similarity coefficient (SDSC) metric\cite{nikolov2018deep}, which was shown to be more representative of any additional effort needed for clinical adaptation\cite{Vassen2020} than the more commonly used geometric metrics like DSC and Hausdroff distances. For completeness, we also report the DSC and Hausdroff distance metric at 95th percentile (HD95) as recommended in prior works\cite{menze2015multimodal}. 

The surface DSC metric emphasizes the incorrect segmentations on the boundary, as this is where the edits are most likely to be performed by clinicians for clinical acceptance. It is computed as:
\begin{equation}
		\begin{split}
		D_{i,j}^{(\tau)}=\frac{\mid S_{i}\cap B_{j}^{(\tau)}  \mid + \mid S_{j}\cap B_{i}^{(\tau)}  \mid }{\mid S_{i} \mid + \mid S_{j} \mid}
		\end{split} 
		\label{eqn:Surface DSC metric}
		\end{equation}
where $B_{i}^{(\tau)}$ $\subset$ $R^{3}$ is a border region of the segmented surface $S_{i}$. The tolerance threshold $\tau =$ 4.38$mm$ was computed using the standard deviation of the HD95 distances of 8 segmentations performed by two different experts blinded to each other.

Statistical comparisons between the various methods was performed to assess the difference between the CMEDL vs. other approaches using paired Wilcoxon two-sided tests using the DSC accuracy measure. Adjustments for multiple comparisons were performed using Holm-Bonferroni method.


			\subsection{\textcolor{black}{Comparison of CMEDL and CBCT segmentation accuracy}}	
\begin{figure*}
				\begin{center}	\includegraphics[width=0.7\columnwidth,scale=0.5]{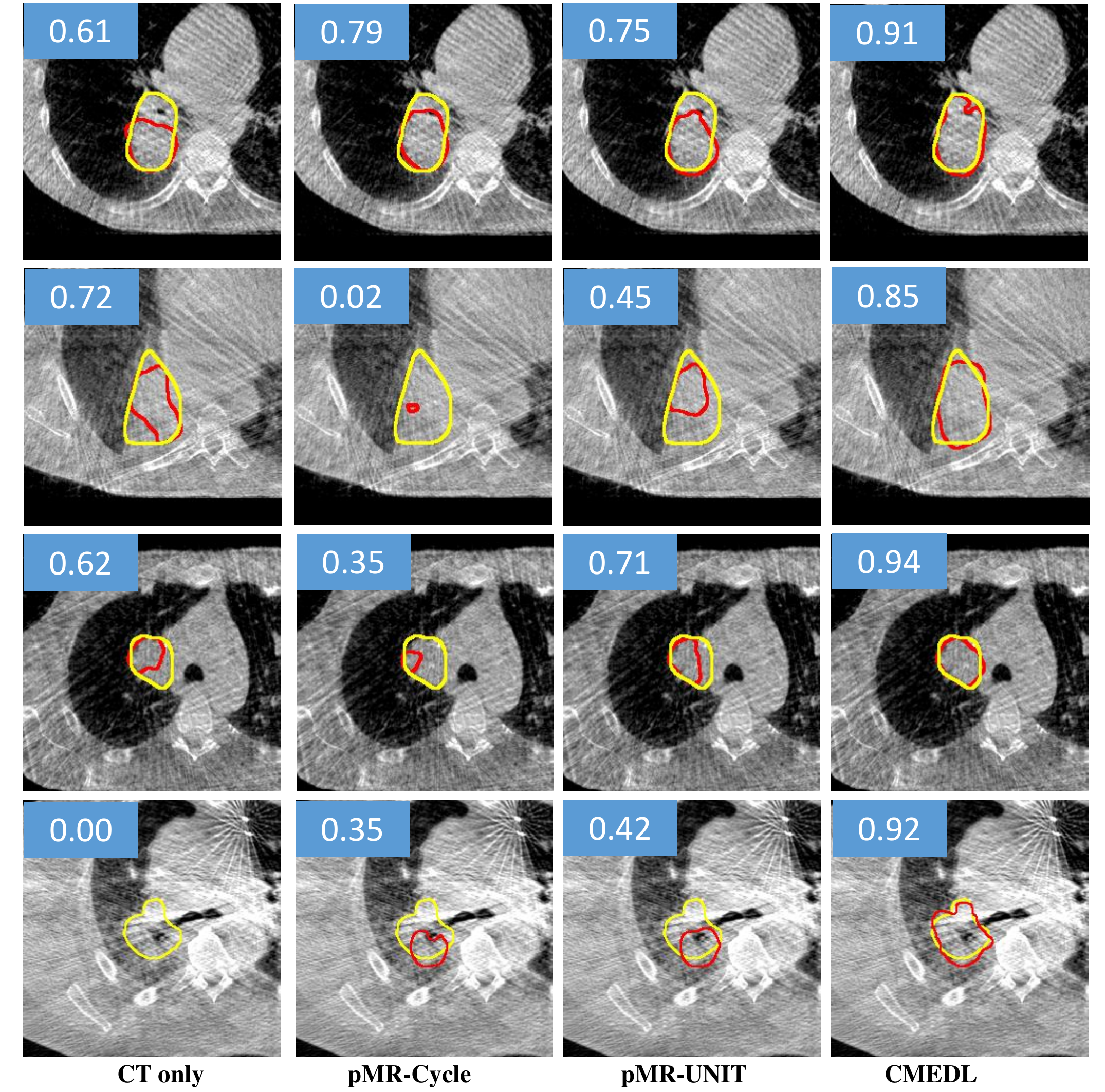}
					\vspace{-0.05cm}\setlength{\belowcaptionskip}{-0.4cm}\setlength{\abovecaptionskip}{0.08cm}\caption{\label{fig:CBCT_Seg} \small Segmentation of representative CBCT images from test set using the CMEDL approach compared with CBCT-only segmentation. Both methods used U-net as a segmentation architecture. Yellow contour indicates the manual segmentation while the red contour indicates the algorithm segmentation. The \textcolor{black}{slice-wise} DSC score of each example is also shown in the blue rectangle area. }
					
				\end{center}
			\end{figure*}

			Table. \ref{tab:CBCT_result_Unet} shows the segmentation accuracies on the validation and test sets produced by various methods for the Unet network. The accuracies for DenseFCN method are also shown in Table.\ref{tab:CBCT_result_Dense}. CMEDL and pMRI-CMEDL methods were significantly more accurate than CBCT only networks using both Unet ($P <0.001$ using DSC and SDSC) and DenseFCN ($P<0.001$ using DSC and SDSC) networks. These two methods were also more accurate than also the networks using pMRI computed from UNIT and cycleGAN networks trained separately from the segmentation network. \textcolor{black}{The approach combining pMRI as a separate channel with the CBCT in the input was the least accurate compared to all other methods indicating that just adding a pMRI as a second channel doesn't contribute to a higher accuracy, possibly as the features from the two modalities are averaged early on. Finally, the CMEDL approach also outperformed a 3DUnet network, underscoring the importance of extracting better features for differentiating target from background than incorporating information from the slices.} Fig.~\ref{fig:CBCT_Seg} shows example segmentations generated on the CBCT images by using the CBCT only, pMRI-Cycle, pMRI-UNIT, and CMEDL methods using Unet network. 
			
			
\subsection{Pseudo MRI translation accuracy}			
\begin{figure*}
				\begin{center}	\includegraphics[width=0.7\columnwidth,scale=0.5]{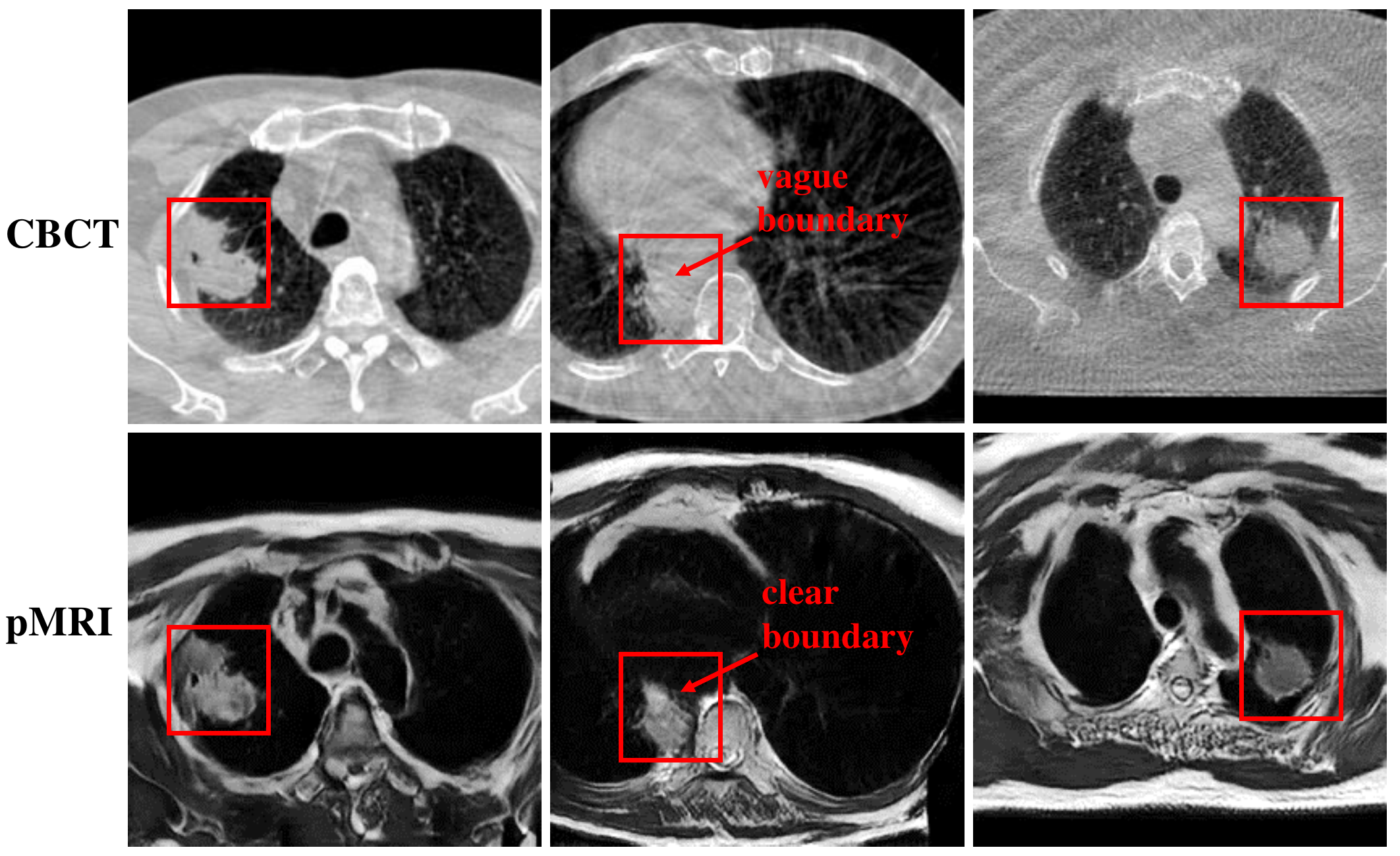}
					\vspace{-0.05cm}\setlength{\belowcaptionskip}{-0.4cm}\setlength{\abovecaptionskip}{0.08cm}\caption{\label{fig:CBCT_Translation} \small Representative examples of pMRI generated from CBCT images using the CMEDL method. The tumor region is enclosed in the red contour.}
					
				\end{center}
			\end{figure*}

We also measured the accuracy of translating the CBCT into pseudo MRI images using the CMEDL, CycleGAN and UNIT methods using Kullback-Leibler (KL) divergence measure. The CMEDL approach resulted in the most accurate translation with the lowest KL divergence of 0.082 when compared with 0.42 for CycleGAN and 0.29 for the UNIT method. Fig.~\ref{fig:CBCT_Translation} shows representative examples from the test set of translated pMRI images produced using the CMEDL method. As seen, tumor and various structures in the MRI are clearly visualized with clear boundary seen between tumor and central structures compared to the corresponding CBCT image. 
			
	        \begin{table*} 
				\centering{\caption{Segmentation accuracy for CBCT dataset using the various approaches using Unet networks.}
					\label{tab:CBCT_result_Unet} 
					\centering
					\scriptsize
					\begin{tabular}{|c|c|c|c|c|c|c|c|} 
						\hline 
						
						\hline 
						{\multirow{2}{*}{Net}}&\multicolumn{1}{|c|}{ }& \multicolumn{3}{|c|}{Validation (N = 20)} & \multicolumn{3}{|c|}{Test (N = 38)}\\ 
						\cline{2-8}
						{}&{  Method  }  & {  DSC  } & { Surface DSC } & {  HD95 $mm$ }& {  DSC  }& {  Surface DSC} & {  HD95 $mm$  }\\
						\hline 
						
						\hline 
						{\multirow{5}{*}{\rotatebox[origin=c]{90}{Unet}}}\multirow{1}{*}&{CBCT only } & { 0.64$\pm$0.21   }& { 0.72$\pm$0.11 } & { 16.19$\pm$11.55   }& { 0.62$\pm0.15$ }& {0.69$\pm$0.11  }   & { 21.70$\pm16.34$ }\\
						&\multirow{1}{*}{pMRI-Cycle } & { 0.65$\pm$0.20}& { 0.72$\pm$0.11  } & { 11.30$\pm$7.64  }& {0.66$\pm$0.12 }& { 0.74$\pm$0.13 } & { 14.43$\pm$12.22}\\
						&\multirow{1}{*}{pMRI-UNIT } & {0.66$\pm$0.19 }& {0.71$\pm$0.10 } & { 10.55$\pm$6.98 }& { 0.66$\pm$0.11 }& { 0.76$\pm$0.13 } & {14.56$\pm$11.80 }\\
						&\multirow{1}{*}{\textcolor{black}{3DUnet}} & {0.66$\pm$0.20 }& {0.70$\pm$0.21 } & { 16.33$\pm$13.07 }& { 0.60$\pm$0.21 }& { 0.72$\pm$0.20 } & {15.01$\pm$12.98 }\\
						&\multirow{1}{*}{\textcolor{black}{CT+pMRI}  } & {0.64$\pm$0.16 }& {0.64$\pm$0.13 } & { 20.01$\pm$8.09 }& { 0.61$\pm$0.17 }& { 0.67$\pm$0.15 } & {19.67$\pm$15.52 }\\						
						\cline{2-8}
						
						\cline{2-8}
						
						&\multirow{1}{*}{pMRI-CMEDL } & {0.74$\pm$0.18}& {0.80$\pm$0.09  } & { 8.87$\pm$8.98   }& { 0.69$\pm$0.10 }& { 0.79$\pm$0.11 } & {10.42$\pm$10.58 }\\
						&\multirow{1}{*}{CMEDL} & {0.73$\pm$0.18}& {0.80$\pm$0.09  } & { 6.39$\pm$6.27   }& {  0.73$\pm$0.10 }& { 0.83$\pm$0.08 } & { 7.69$\pm$7.86 }\\
						\hline 
						
						\hline

					\end{tabular}
				}
			\end{table*}

	        \begin{table*} 
				\centering{\caption{Segmentation accuracy for CBCT dataset using the various approaches using the DenseFCN network.}
					\label{tab:CBCT_result_Dense} 
					\centering
					\scriptsize
					\begin{tabular}{|c|c|c|c|c|c|c|c|} 
						\hline
						
						\hline 
						{\multirow{2}{*}{Net}}&\multicolumn{1}{|c|}{ }& \multicolumn{3}{|c|}{Validation (N = 20)} & \multicolumn{3}{|c|}{Test (N = 38)}\\ 
						\cline{2-8}
						{}&{  Method  }  & {  DSC  } & { Surface DSC } & {  HD95 $mm$ }& {  DSC  }& {  Surface DSC} & {  HD95 $mm$  }\\
						\hline 
						
						\hline 
							
						{\multirow{5}{*}{\rotatebox[origin=c]{90}{DenseFCN}}}\multirow{1}{*}&{CBCT only } & { 0.63$\pm$0.18   }& { 0.59$\pm$0.12 } & { 22.88$\pm$12.91   }& { 0.58$\pm0.14$ }& {0.66$\pm$0.15  }   & { 22.15$\pm17.19$ }\\
						&\multirow{1}{*}{pMRI-Cycle } & { 0.65$\pm$0.13}& { 0.63$\pm$0.12  } & { 14.83$\pm$11.83  }& {0.57$\pm$0.14 }& { 0.66$\pm$0.18 } & { 21.82$\pm$15.97}\\
						&\multirow{1}{*}{pMRI-UNIT } & {0.66$\pm$0.14 }& {0.65$\pm$0.15 } & { 20.98$\pm$15.60 }& { 0.64$\pm$0.15 }& { 0.62$\pm$0.17 } & {25.28$\pm$15.94 }\\
						\cline{2-8}
						
						\cline{2-8}
						&\multirow{1}{*}{pMRI-CMEDL } & {0.69$\pm$0.18 }& {0.69$\pm$0.10 } & { 17.60$\pm$9.25 }& { 0.68$\pm$0.12  }& { 0.73$\pm$0.13 } & {14.46$\pm$11.87 }\\
						&\multirow{1}{*}{CMEDL } & { 0.69$\pm$0.17}& {0.70$\pm$0.11 } & {11.43$\pm$6.91   }& { 0.72$\pm$0.13  }& {0.75$\pm$0.13  } & { 11.42$\pm$9.87   }\\
						\hline
						
						\hline 
					\end{tabular}
				}
			\end{table*}
			
			\begin{figure*}
				\begin{center}	\includegraphics[width=1.0\columnwidth,scale=0.5]{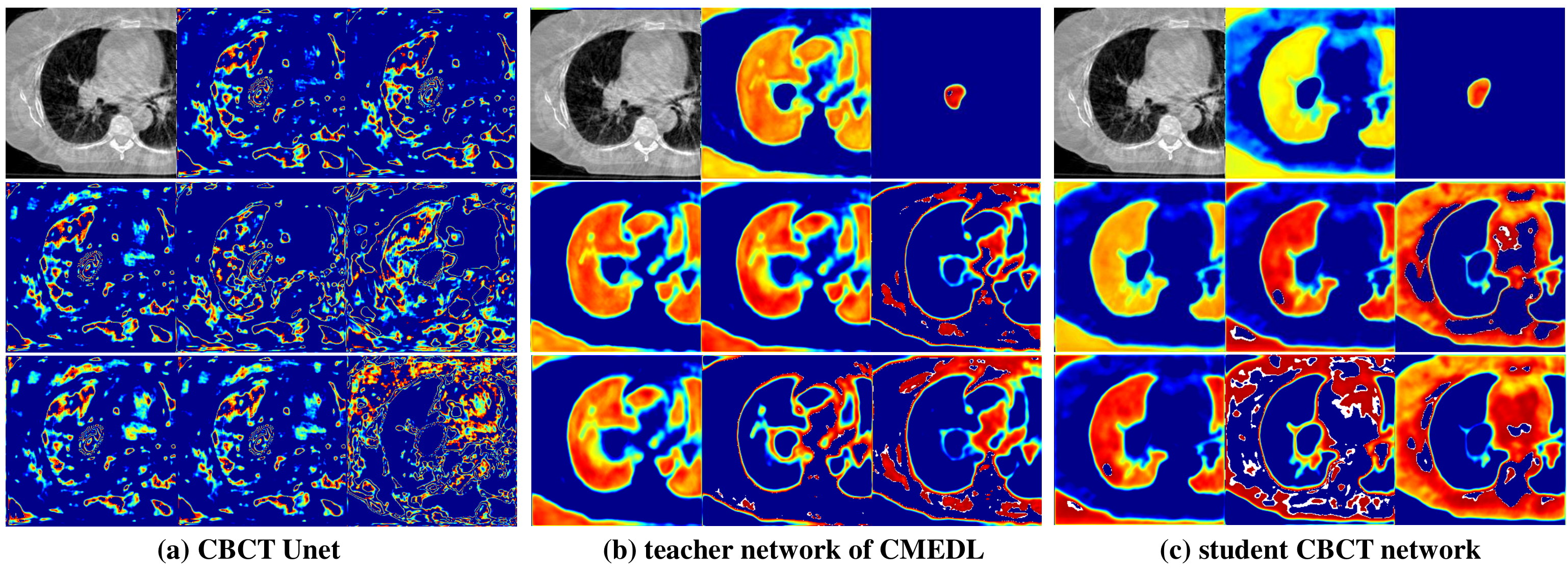}
					\vspace{-0.05cm}\setlength{\belowcaptionskip}{-0.4cm}\setlength{\abovecaptionskip}{0.08cm}\caption{\label{fig:feature_map} \small Feature activaton maps computed from (a) CBCT Unet, (b) teacher network of CMEDL, and (c) student CBCT network.}
					
				\end{center}
			\end{figure*} 
		
			\begin{figure*}
				\begin{center}	\includegraphics[width=0.6\columnwidth,scale=0.5]{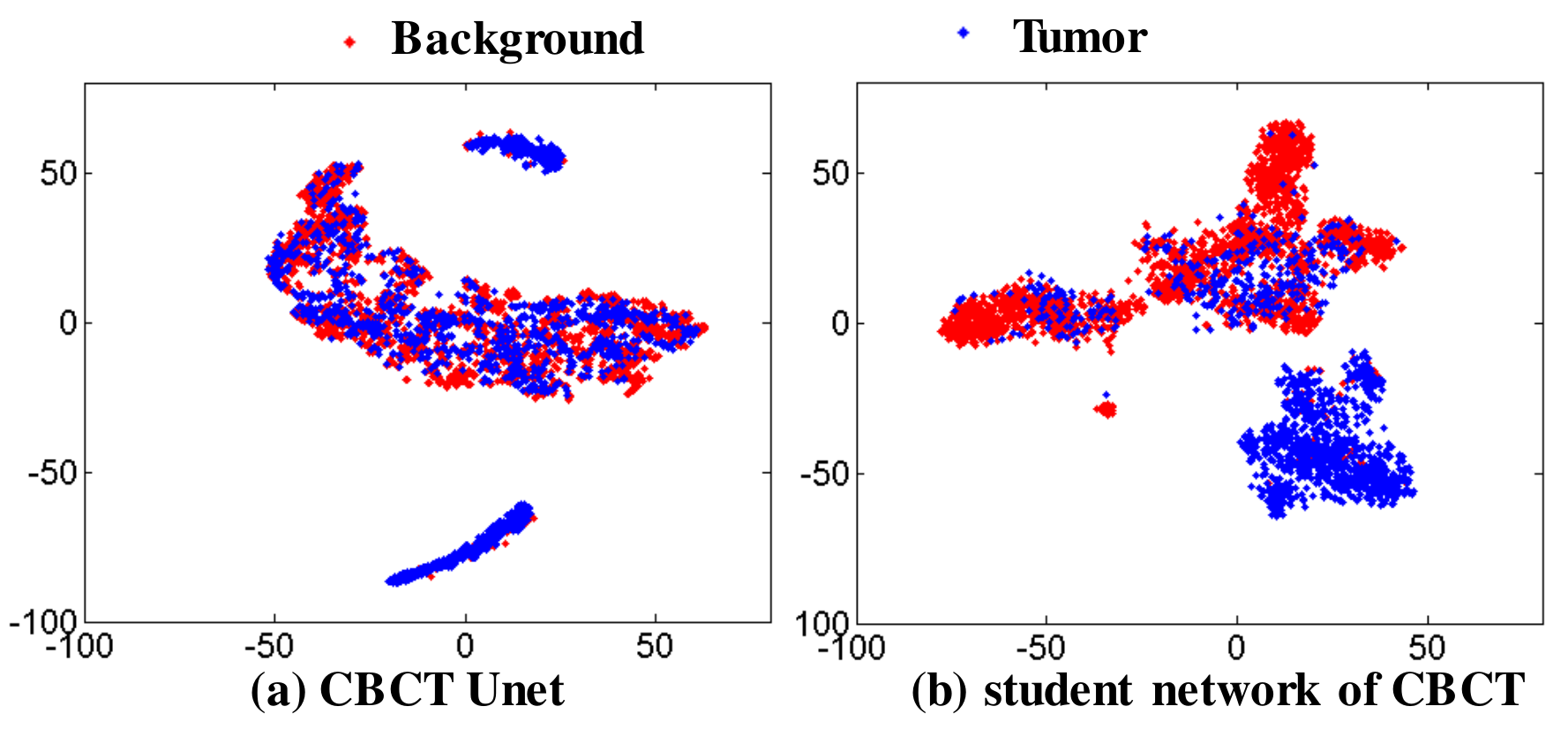}
					\vspace{-0.05cm}\setlength{\belowcaptionskip}{-0.4cm}\setlength{\abovecaptionskip}{0.08cm}\caption{\label{fig:TSNE} \small \textcolor{black}{TSNE map of Feature activaton maps computed from (a) CBCT Unet, (b) CMEDL-CBCT Unet.}}
					
				\end{center}
			\end{figure*} 

			\begin{figure*}
				\begin{center}	\includegraphics[width=1.0\columnwidth,scale=0.5]{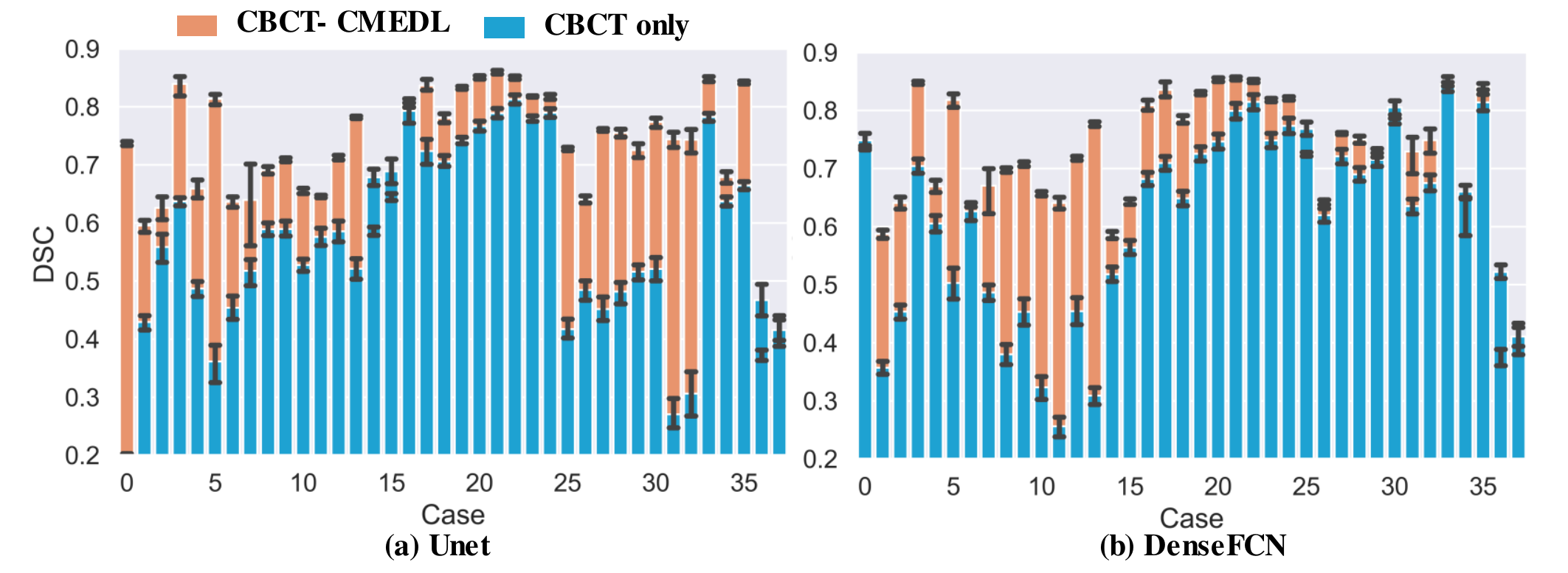}
					\vspace{-0.05cm}\setlength{\belowcaptionskip}{-0.4cm}\setlength{\abovecaptionskip}{0.08cm}\caption{\label{fig:sensitivity} \small \textcolor{black}{Sensitivity analysis using test time dropout for CMEDL vs. CBCT only networks: (a) Unet, (b) DenseFCN. The segmentation variabilities per each test case for these methods are also shown.}}
					
				\end{center}
			\end{figure*}

\subsection{Sensitivity analysis}
\textcolor{black}{We evaluated the robustness of the networks' segmentation by introducing noise into the learned models. This was done by performing random dropout of the weights in the last two network layers during testing. Test-time dropout was performed by keeping the dropout rate at 0.5 and randomly zeroing out the learned weights. Dropout was run 10 times for each data and the average of each case is shown in Figure \ref{fig:sensitivity}. Both of the CMEDL networks resulted in lower variabilities in the segmentation as shown in Fig.~\ref{fig:sensitivity} for each test case. The \textcolor{black}{mean standard deviation (mSD) was 0.016 using DSC for CMEDL-Unet, 0.016 using DSC for CMEDL-DenseFCN. Whereas the mSD was higher for both networks trained using only CBCT data with a 0.024 mSD for DSC of Unet and 0.027 mSD of DSC for denseFCN.}}

\subsection{Separation of target from background using feature maps extracted from CMEDL and CBCT only networks}
Finally, we \textcolor{black}{studied} how the cross-modality anatomical information from MRI lead to segmentation improvement on CBCT. Fig.\ref{fig:feature_map} shows an example case with eight randomly chosen feature maps selected from the last layer (with size of 256$\times$256$\times$64) of Unet network trained with only CBCT  (Fig.\ref{fig:feature_map}(a)) and the corresponding feature maps and the same case when computed from a CMEDL-Unet (Fig.\ref{fig:feature_map}(c)). For reference, the feature maps computed in the same layer for the teacher MRI network that used the translated pMRI are also shown (Fig.\ref{fig:feature_map}(b)). As seen, the feature maps extracted using the CBCT only method are less effective in differentiating the tumor regions from the background parenchyma when compared to the CMEDL CBCT network. \textcolor{black}{The feature maps extracted from CBCT using the student network are highly similar to the feature maps extracted from the corresponding pMRI images extracted from the teacher network, indicating sufficient knowledge distillation between the teacher and student networks. As a result, the CBCT student network extracted features that clearly distinguished the tumor from the background even when it is fed only CBCT images. This in turn produced more accurate segmentation than a CBCT only network trained without CMEDL approach.} 

\textcolor{black}{Fig.\ref{fig:TSNE} shows the result of unsupervised clustering the feature maps from the last layer (256$\times$256$\times$64) for the CMEDL and CBCT only Unet networks for all \textcolor{black}{the 38 test cases.} Note that during test only the student CBCT Unet network of the CMEDL method is used. The input to the clustering method consisted of randomly selected set of pixel features chosen from inside a \textcolor{black}{160$\times$160} pixels region of interest enclosing the tumor in each slice containing the tumor, resulting in a total of 200,000 pixels. The number of pixels corresponding to target and background was balanced. Clustering was done by using the t-distributed stochastic neighbor embedding (t-SNE)\cite{van2008visualizing} method using the Matlab. Briefly, the t-SNE method computes an unsupervised clustering of high dimensional data by computing a high dimensional and low-dimensional embedding of the data as probability distributions. Gradient descent is used to minimize the Kullback-Leibler divergence between the two distributions either until convergence or a maximum number of iterations. The clustering parameters, namely perplexity, which is related to the number of nearest neighbors was set of 60 and the number of gradient descent iterations to 1000. As seen in Fig.~\ref{fig:TSNE}, the features extracted from the CBCT Unet network trained using CMEDL are better able to distinguish the tumor from background pixels than the features extracted from a CBCT Unet alone.}

\section{Discussion}
We introduced a new approach to leverage higher contrast information from more informative MRI modality to improve CBCT segmentation. Our approach uses unpaired sets of MRI and CBCT images from different sets of patients and learns to extract informative features on CBCT that help to obtain a good lung tumor segmentation, even for some centrally located tumors. Our approach shows a clear improvement over CBCT only segmentation \textcolor{black}{using both 2D and 3D methods} as well as pMRI based segmentation when the pMRI is generated using UDA networks trained independent of the CMEDL framework. \textcolor{black}{Importantly, our results also showed that the CMEDL approach enables the CBCT network to extract more informative features that better distinguish tumor from background. This arises by guiding the CBCT student network to extract features that match the statistical distribution of the teacher MRI network.}

To our best knowledge, this is one of the first works to address the problem of fully automatic lung tumor segmentation on CBCT images. Prior work on CBCT used semi-automated segmentation\citep{Veduruparthi2018}, and the CBCT deep learning methods were applied to segment pelvic normal organs\citep{Fu2020,Lei2020}. \textcolor{black}{A clear difference of our approach from the deep learning methods is the use of pMRI as side information during training to guide feature extraction from the CBCT network. The method in Fu et.al\citep{Fu2020} combined pMRI with the CBCT features even during testing as a late fusion network while\cite{Lei2020} performed segmentation from pMRI image patches using attention gated Unet. The main advantage of our method is that only a lightweight CBCT segmentation network is needed for testing and the requirements for pMRI accuracy are less stringent than the methods\cite{Fu2020,Lei2020}, which use pMRI as an input modality.} 

Our results showed that the MRI information educed on CBCT is most useful when used as hints through cross-modality distillation. \textcolor{black}{This is because the teacher network provides additional regularization to guide the extraction of CBCT features that yeild the best possible segmentation performance during training. This regularization is accomplished by matching the features computed from CBCT with the features computed from the corresponding pMRI images. On the other hand, a CBCT network that does not use such constraints may result in a local minimum after training, but this is not guaranteed to extract features that better distinguish tumor from background as shown in our results.} Similarly, the frameworks that used the synthesized pMRIs directly for segmentation were also less accurate than the CMEDL CBCT network. In the extreme case using pMRI as a secondary input channel with CBCT led to degradation in performance. This is because distillation learning itself only provides additional regularization to constrain the set of features extracted by the student CBCT network, which does not require as accurate pMRI translation as would be needed when using pMRI itself as input for the segmentation network. On the other hand, accurate I2I translation is more important when using the translated image for segmentation. As shown, both pMRI-Cycle and pMRI-UNIT were comparable in performance. On the other hand, the pMRI-CMEDL approach, which uses side information from CBCT student network to also constrain the I2I translation improves the accuracy of the teacher network for segmenting CBCT. \textcolor{black}{However, all these methods were less accurate than when using the student CMEDL CBCT network for segmentation.}

As opposed to standard distillation methods that used a pre-trained teacher network as done in \cite{chen2017learning, romero2014fitnets, gupta2016cross}, which required paired image sets, ours is the first, to our best knowledge, that works with completely unrelated set of images from widely different imaging modalities. Removing the constraint of paired image sets makes our approach more practical for medical applications, including new image-guided cancer treatments. 

\textcolor{black}{As a limitation, our approach used a modest number of CBCT images for training and testing. Addition of more training sets would likely improve performance even more. Similarly, use of a 3D architecture instead of a 2D architecture could enable obtaining more accurate volume segmentations. Also, testing on  multi-institutional datasets with different imaging acquisitions is essential to establish the generality of the developed approach and is work for future.} To summarize, to our best knowledge, this is the first approach to tackle the problem of CBCT lung tumor segmentation.    
			
\section{Conclusion}
We introduced a novel cross-modality educed distillation learning approach for segmenting lung tumors on cone-beam CT images. Our approach uses unpaired MRI and CBCT image sets to constrain the features extracted on CBCT to improve inference and segmentation performance. Our approach implemented on two different segmentation networks showed clear performance improvements over CBCT only methods. Evaluation on much larger datasets is essential to assess potential for clinical translation.  
		
\section{Acknowledgements}
This research was supported by NCI [grant number R01-CA198121]. It was also partially supported through the NIH/NCI Cancer Center Support Grant [grant number P30 CA008748] who had no involvement in the study design; the collection, analysis and interpretation of data; the writing of the report; and the decision to submit the article for publication.   
		
\section{References}
\bibliographystyle{medphy}
\bibliography{refs}



\end{document}